\documentclass[aip,jcp,reprint,superscriptaddress]{revtex4-1}
\usepackage{graphicx}
\usepackage{amsmath}
\usepackage{physics}
\usepackage{braket}
\usepackage{amssymb}
\usepackage{mathtools}
\usepackage{multirow}
\usepackage[colorlinks,citecolor=blue,urlcolor=black]{hyperref}

\def\ben{\begin{equation}}
\def\een{\end{equation}}
\def\mr{\mathbf{r}}
\def\mg{\mathbf{G}}
\def\mq{\mathbf{q}}

\begin{document}

\title{Dielectric embedding $GW$ for weakly coupled molecule-metal interfaces}

\author{Zhen-Fei Liu}
\affiliation{Department of Chemistry, Wayne State University, Detroit, MI 48202, USA}

\date{\today}

\begin{abstract}
Molecule-metal interfaces have a broad range of applications in nanoscale materials science. Accurate characterization of their electronic structures from first-principles is key in understanding material and device properties. The $GW$ approach within many-body perturbation theory is the state-of-the-art and can in principle yield accurate quasiparticle energy levels and interfacial level alignments that are in quantitative agreement with experiments. However, the interfaces are large heterogeneous systems that are currently challenging for first-principles $GW$ calculations. In this work, we develop a $GW$-based dielectric embedding approach for molecule-metal interfaces, significantly reducing the computational cost of direct $GW$ without sacrificing the accuracy. To be specific, we perform explicit $GW$ calculations only in the simulation cell of the molecular adsorbate, in which the dielectric effect of the metallic substrate is embedded. This is made possible via a real-space truncation of the substrate polarizability and the use of the interface plasma frequency in the adsorbate $GW$ calculation. Here, we focus on the level alignment at weakly coupled molecule-metal interfaces, i.e., the energy difference between a molecular frontier orbital resonance and the substrate Fermi level. We demonstrate our method and assess a few $GW$-based approximations using two well-studied systems, benzene adsorbed on the Al (111) and on the graphite (0001) surfaces.
\end{abstract}

\maketitle

\section{Introduction}


Interfaces formed by molecules adsorbed on metallic substrates are central to a number of applications in nanoscience \cite{KUW13}, such as photovoltaics \cite{HBSK10} and optoelectronics \cite{LTOS16}. Unlike bulk materials, interfaces are intrinsically heterogeneous, where two different materials interact and yield properties that are not present in the bulk. For instance, interface dipoles are formed, as a result of bonding and the change in electron density compared to the isolated molecule/substrate limit \cite{L71}. From an electronic structure theory perspective, these effects are mainly due to one-body interactions at the interface, in the sense that electron-electron correlation does not play a dominant role. Therefore, typical local and semi-local functionals in density functional theory (DFT) \cite{KS65} can yield accurate results for aforementioned properties \cite{CGCC02}. However, there are also other interfacial properties that explicitly depend on the electron correlation across the interface, with one prominent example being the energy level alignment \cite{NHL06} - how molecular frontier resonances are aligned with the Fermi level of the substrate. This is because the substrate serves as a dielectric media that effectively screens the Coulomb interaction within the molecule, which renormalizes the gap between frontier orbitals of the adsorbate compared to the isolated molecule limit. This is the so-called ``image-charge effect'' \cite{LK73,NHL06,TR09} that requires beyond-DFT treatments \cite{QVCL07}. Physically, the level alignment determines the charge-transfer barrier across the interface \cite{HRKH98} and hence affects the interfacial dynamics such as charge transport \cite{DKCV10}. Accurate characterization of the energy level alignment is then crucial for fundamental understanding of material and device properties.


A formally rigorous approach to calculate the level alignment is many-body perturbation theory (MBPT), as the energy levels of interest at the interface are in fact quasiparticle energy levels, i.e., particle-like charged excitation levels in an interacting system, conceptually different from the Kohn-Sham (KS) energy levels in DFT. Quantitatively, local and semi-local functionals often underestimate the level alignment by about 1 eV or more \cite{NHL06,QVCL07}, leading to large errors in the prediction of interfacial electronic structure and dynamics \cite{DKCV10}. Within the framework of MBPT, the $GW$ approximation ($G$ is the Green's function and $W$ is the screened Coulomb interaction) to the self-energy is often employed, as the first-order perturbation correction to the KS eigenvalues \cite{H65,HL86}. With significant advancements in algorithms \cite{BerkeleyGW,GG15} and computer hardwares, $GW$ has been proved very successful for small systems such as molecules \cite{SCSR15} and bulk solids \cite{MC13}. Still, with the large system size and the heterogeneity, molecule-substrate interfaces remain computationally challenging for $GW$.


A historically useful approach to treat large systems is the embedding theory \cite{SC16}, where a small subsystem of interest is typically treated explicitly or with higher accuracy and the effect of its environment is taken into account in an implicit and efficient manner with typically lower accuracy. Embedding theory has been successfully applied to wavefunction \cite{BGMM13,LHC14} and density functional \cite{GWC99,MSGM12} based approaches, mostly for the purpose of computing total energy related properties. Embedding formalisms are particularly suitable for cases in which the physical focus is on a small portion of the entire system and the system/environment are naturally partitioned and defined, such as adsorbate on substrates \cite{TBCI96} and molecules in solvent \cite{JNJV06}. Weakly coupled molecule-metal interfaces that we study in this work are such cases: the nature of the frontier orbitals of the adsorbate typically carries more chemical meanings for material and device properties than those of the substrate \cite{GBEE16,RPWT16}, as the substrate is often used to provide support to and tune the property of the adsorbed molecule. Therefore, it is meaningful to leverage the idea of embedding for efficient calculations of the quasiparticle electronic structure at weakly coupled molecule-metal interfaces, which is the primary goal in this work.


For weakly coupled molecule-metal interfaces, orbital hybridizations at the interface are negligible, and the metallic substrate mainly provides a dielectric environment for the adsorbate \cite{UBSJ14,ALT15,QJL17,XCQ19,LJLN19}, therefore we term our approach dielectric embedding $GW$. This embedding idea was first proposed in Ref. \citenum{UBSJ14}, in the context of MoSe$_2$-bilayer graphene interface. However, in Ref. \citenum{UBSJ14}, the local-field effect is neglected, and the simulation cell of the adsorbate shares the same dimension along the $z$ (surface normal) direction as the interface. Here, we move forward by considering the physical coverage of the molecular adsorbate hence capturing the local-field effect exactly, and embed the substrate dielectric effect into a simulation cell of a much smaller size along $z$ (see Fig. \ref{cartoon}), further reducing the computational cost. This is achieved via a real-space truncation of the substrate KS polarizability, a reverse procedure of the real-space mapping that we developed in Ref. \citenum{LJLN19}. Additionally, In prior works such as Refs. \citenum{UBSJ14,CTQ17,QJL17,LJLN19}, the difference between self-energies calculated using the Hybertsen-Louie generalized plasmon-pole (GPP) model \cite{HL86} and the static Coulomb-hole-screened-exchange (COHSEX) approximation \cite{JDSC14} is assumed to be the same for the isolated molecule and for the adsorbate within the interface, to accommodate the difference between plasmon poles of the molecule and the interface. Here, we take a different approach by directly using the plasma frequency of the interface in the adsorbate self-energy calculation using the GPP model, avoiding $GW$ calculations of the interface as well as the intrinsic inaccuracies of COHSEX. As a proof of concept, we demonstrate the feasibility of our approach using well-studied while experimentally relevant systems \cite{DMBN00,NHL06}, benzene adsorbed on the Al (111) and on the graphite (0001) surfaces. We assess our approach by comparing results obtained from different levels of approximation within the $GW$ framework.

This paper is structured as follows. In Sec. \ref{sec:metho} we describe our method in detail, especially the dielectric embedding technique and the use of the interface plasma frequency in adsorbate calculations. In Sec. \ref{sec:res} we discuss our results by comparing level alignments calculated using different levels of theory. Sec. \ref{sec:dis} is devoted to discussions and we conclude in Sec. \ref{sec:conc}.

\section{Methodology}
\label{sec:metho}

Our goal is to deduce the energy level alignment for the interface based on a dielectric embedding $GW$ calculation of the molecular adsorbate alone. There are three issues to address: (1) the substrate dielectric effect needs to be embedded into the adsorbate simulation cell; (2) in a GPP calculation, the plasmon poles of the adsorbate are in general different from those of the interface, which needs to be taken into account; and (3) in the self-energy calculation of the interface, some matrix elements of $\Sigma$ (self-energy) involve the screened Coulomb interaction between orbitals localized on the adsorbate and orbitals localized on the substrate. However, in an embedding adsorbate-only calculation, there are no substrate orbitals. In this work, we neglect these contributions to the self-energy for weakly coupled molecule-metal interfaces, by assuming no spatial overlap between adsorbate and substrate orbitals. This is justified by the accuracy of the results. We address the first two issues in this section.

\subsection{Embedding the Dielectric Effect of the Substrate into the Simulation Cell of the Adsorbate}
\label{sec:embed}

One of the major bottlenecks for large-scale $GW$ calculations is the non-interacting KS polarizability $\chi^0$ within the random-phase approximation (RPA), which has a formal scaling of ${\cal O}(N^4)$ with $N$ the number of KS orbitals \cite{BerkeleyGW}. It is not only CPU-intensive, but also memory-intensive, given the fact that typically thousands of unoccupied orbitals are required to converge the results. In Ref. \citenum{LJLN19}, we developed an approach to break down this $\chi^0$ calculation into smaller and affordable ones for the building blocks of the interface: the periodic molecular layer and the periodic substrate. Since the $\chi^0$ is additive in real space for weakly coupled interfaces, here we consider the sum of the adsorbate $\chi^0$ and the substrate $\chi^0$ around the region of the molecular adsorbate:
\ben
\tilde{\chi}^0_{\rm tot} \approx \tilde{\chi}^0_{\rm sub} + \chi^0_{\rm mol}.
\label{eq:sum}
\een
This approximation holds for weakly coupled molecule-substrate interfaces, as shown in Refs. \citenum{LJLN19,XCQ19}. Here all quantities are defined in the simulation cell of the adsorbate. $\tilde{\chi}^0_{\rm tot}$ is the effective polarizability that the adsorbate would ``feel'' in the interface, $\tilde{\chi}^0_{\rm sub}$ is the polarizability of the substrate truncated into the adsorbate simulation cell, and $\chi^0_{\rm mol}$ is the polarizability of the periodic molecular layer.

\begin{figure}[htbp]
\begin{center}
\includegraphics[width=3.5in]{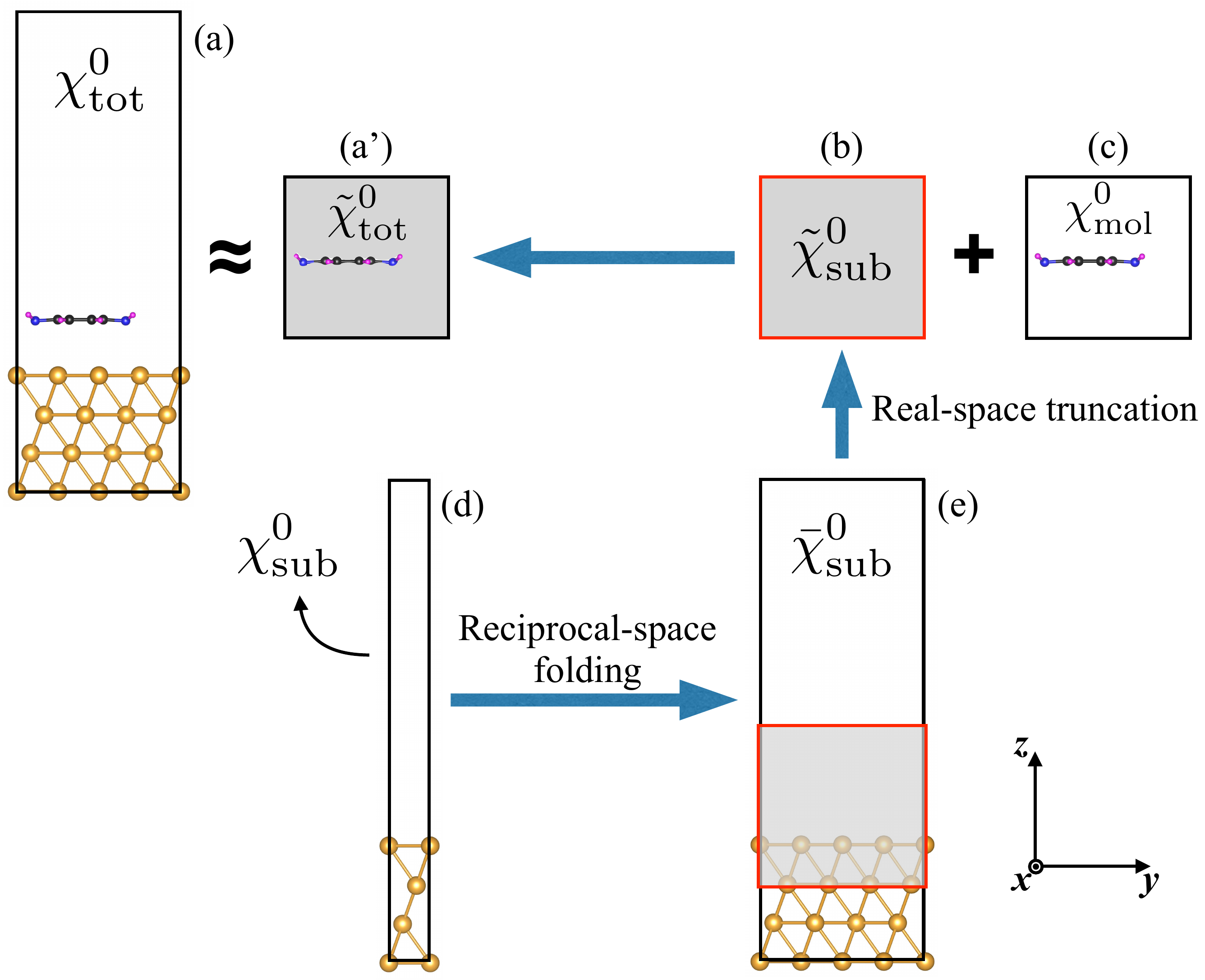}
\caption{An overview and the workflow of the dielectric embedding $GW$ approach. (a') is the simulation cell where $GW$ self-energies of the adsorbate frontier orbitals are calculated. Here the molecule ``feels'' a dielectric environment of the substrate, denoted by the gray background. (a') is an embedded version of the full interface (a), which is more computationally demanding and is what we aim to avoid in this work. (c) and (d) are the two systems for which $\chi^0$ needs to be calculated using the standard RPA formula, which is much more affordable than (a). For other details, see text.}
\label{cartoon}
\end{center}
\end{figure}

Fig. \ref{cartoon} shows schematically an overview and the workflow of our approach. In (a), the total interface system containing both the adsorbed molecule and the substrate is shown, with KS polarizability $\chi^0_{\rm tot}$. The molecular resonance is determined as the peak in the projected density of states (PDOS), with its $GW$ self-energy calculated using $\braket{\phi_i^{\rm tot}|\Sigma[G^{\rm tot}W^{\rm tot}]|\phi_i^{\rm tot}}$. Here, $\ket{\phi_i^{\rm tot}}$ is an orbital of the interface at the PDOS peak that resembles a molecular frontier orbital. The self-energy $\Sigma$ depends on $G^{\rm tot}$ and $W^{\rm tot}$, where $G^{\rm tot}$ is the Green's function of the interface and $W^{\rm tot}$ is the screened interaction, defined as $W^{\rm tot}=\epsilon^{-1}v$ with $v$ the bare Coulomb interaction operator. The inverse dielectric function, $\epsilon^{-1}$ is calculated using \cite{BerkeleyGW} $\epsilon^{-1}=[1-v\chi^0_{\rm tot}]^{-1}$. The main idea in our approach is to write
\ben
\begin{split}
\braket{\phi_i^{\rm tot}|\Sigma[G^{\rm tot}W^{\rm tot}]|\phi_i^{\rm tot}} & \approx \braket{\phi_i^{\rm mol}|\Sigma[G^{\rm tot}W^{\rm tot}]|\phi_i^{\rm mol}} \\
& \approx \braket{\phi_i^{\rm mol}|\Sigma[G^{\rm mol}\tilde{W}^{\rm tot}]|\phi_i^{\rm mol}}.
\end{split}
\label{eq1}
\een
The first line is the so-called ``molecular projection'' approximation developed and used in Refs. \citenum{TDQB11,CTQ17}, which has been shown to be accurate for many weakly coupled molecular resonances on metallic substrates. In the second line of Eq. \eqref{eq1}, we further propose to calculate every quantity in the simulation cell containing just the adsorbate, with an effective screened Coulomb interaction $\tilde{W}^{\rm tot}$ such that the dielectric effect of the substrate is taken into account. This is schematically shown in Fig. \ref{cartoon}(a'), where the gray background shows that the dielectric environment that the molecule ``feels'' is $\tilde{\chi}^0_{\rm tot}$ instead of merely $\chi^0_{\rm mol}$. The tilde means a quantity truncated in real space to a smaller simulation cell. In this way, the dielectric effect of the substrate is ``embedded'' into the simulation cell of the adsorbate, which can be much smaller in size than the interface simulation cell.

We note in passing that in the second line of Eq. \eqref{eq1}, matrix elements involving the screened Coulomb interaction between the molecular orbital $\ket{\phi^{\rm mol}}$ and the interface orbitals that are localized on the substrate (terms that are in $G^{\rm tot}$ but not included in $G^{\rm mol}$) are neglected, compared to the first line of Eq. \eqref{eq1}. Although this approximation needs further investigation for general cases, we argue that the contributions from such terms are small if the spatial overlap between the adsorbate and the substrate orbitals is small. This is justified by the results for the two systems studied in this paper.

As we discussed in Eq. \eqref{eq:sum}, the $\tilde{\chi}^0_{\rm tot}$ in Fig. \ref{cartoon}(a') is further approximated as the sum of the KS polarizabilities of the periodic molecular layer, $\chi^0_{\rm mol}$ in Fig. \ref{cartoon}(c), and that of the substrate truncated to the adsorbate simulation box, $\tilde{\chi}^0_{\rm sub}$ in Fig. \ref{cartoon}(b). The $\chi^0_{\rm mol}$ can be calculated using the standard RPA approach as implemented in \texttt{BerkeleyGW} \cite{BerkeleyGW}. The $\tilde{\chi}^0_{\rm sub}$ is truncated in real space from that of the substrate supercell, $\bar{\chi}^0_{\rm sub}$ in Fig. \ref{cartoon}(e). The dimensions of Fig. \ref{cartoon}(e) and Fig. \ref{cartoon}(c) along the extended directions ($xy$) need to be the same as those in the interface system in Fig. \ref{cartoon}(a), in order to capture the local-field effects exactly. We truncate the $\bar{\chi}^0_{\rm sub}$ in real space to a region that is denoted by the gray area with the red border, whose size needs to be consistent with that of Fig. \ref{cartoon}(c), the isolated periodic molecular layer. We choose a truncation region centered around the adsorbate (as if it is in the interface) along the $z$ direction. The size of Fig. \ref{cartoon}(b)(c) along the $z$ direction needs to be large enough to account for the long-range screened Coulomb interaction across the interface. It practically means that the truncation region should extend to a few Angstroms below the top atomic layer of the substrate if the substrate is metallic, a point to be further elaborated below. This real-space truncation is technically realized via a reverse procedure of the real-space mapping technique that we developed in Ref. \citenum{LJLN19}. Finally, the quantity $\bar{\chi}^0_{\rm sub}$ in Fig. \ref{cartoon}(e) can be efficiently and exactly calculated via a reciprocal-space folding procedure from a unit cell of the substrate, as in Fig. \ref{cartoon}(d), following Refs. \citenum{LJLN19,XCQ19}. This is because with periodic boundary conditions, Fig. \ref{cartoon}(d) and Fig. \ref{cartoon}(e) describe the same physical system and contain the same amount of information in the KS polarizability. The $\chi^0_{\rm sub}$ of the substrate unit cell in Fig. \ref{cartoon}(d) can be calculated using the standard RPA approach as implemented in \texttt{BerkeleyGW} \cite{BerkeleyGW}.

After $\tilde{\chi}^0_{\rm tot}$ is calculated in Fig. \ref{cartoon}(a'), the $\tilde{W}^{\rm tot}$ in Eq. \eqref{eq1} is calculated via $\tilde{W}^{\rm tot}=\tilde{\epsilon}^{-1}v$ with $\tilde{\epsilon}^{-1}=[1-v\tilde{\chi}^0_{\rm tot}]^{-1}$. With the effective dielectric function, the embedding self-energy calculations are performed using the standard approach as implemented in \texttt{BerkeleyGW} \cite{BerkeleyGW}.

\subsection{Using the Plasma Frequency of the Interface in the Embedding Calculation}
\label{sec:plasma}

For a meaningful comparison between the second and the third expressions in Eq. \eqref{eq1}, besides the difference in $G$, another factor needs to be considered: the positions of the plasmon poles are different for the interface and for the adsorbate, when the GPP model is used for the frequency dependence of the dielectric function. The full-frequency calculations are free of this issue, but the computational cost would be roughly $N_f$ times that of a GPP calculation with $N_f$ the number of frequency grids used. We limit our discussions to the Hybertsen-Louie GPP model \cite{HL86} in this work.

To accommodate the difference in the plasmon poles, Ref. \citenum{UBSJ14} assumed that the difference between the self-energies calculated using the GPP and the COHSEX is the same for the interface in the large simulation cell and for the adsorbate in the small simulation cell. However, this approximation is not fully justified for a broad range of systems. Here, we take a different approach, by directly using the plasma frequency of the interface in embedding self-energy calculations of the adsorbate. We start our discussion from the definition of the plasma frequency \cite{HL86,ZTCL89,BerkeleyGW}
\ben
\Omega^2_{\mathbf{G}\mathbf{G}'}(\mathbf{q})=\omega_p^2\frac{(\mq+\mg)\cdot(\mq+\mg')}{|\mq+\mg|^2}\frac{\rho(\mg-\mg')}{\rho(\mathbf{0})}.
\label{gpp}
\een
Here, $\omega_p^2=4\pi \rho(\mathbf{0})e^2/m$ is the classical plasma frequency. Because the volume and the number of electrons are different for the adsorbate simulation cell and for the interface simulation cell, $\rho(\mathbf{0})$ and $\omega_p^2$ are different for the two systems. In this work, in the calculation of $\Omega^2_{\mathbf{G}\mathbf{G}'}(\mathbf{q})$ for the embedded adsorbate, we use the $\omega_p^2$ and the $\rho(\mg-\mg')/\rho(\mathbf{0})$ values from the interface system. In other words, we modify the plasma frequencies in the embedding calculation to those of the interface \cite{note1}.

The GPP mode frequency, the pole appearing in the self-energy expression, is defined as \cite{BerkeleyGW} $\tilde{\omega}^2_{\mathbf{G}\mathbf{G}'}(\mathbf{q})=|\lambda_{\mathbf{G}\mathbf{G}'}(\mathbf{q})|/\cos \phi_{\mathbf{G}\mathbf{G}'}(\mathbf{q})$, with the amplitude and the phase defined by
\ben
|\lambda_{\mathbf{G}\mathbf{G}'}(\mathbf{q})|e^{i\phi_{\mathbf{G}\mathbf{G}'}(\mathbf{q})}=\frac{\Omega^2_{\mathbf{G}\mathbf{G}'}(\mathbf{q})}{\delta_{\mathbf{G}\mathbf{G}'}-\tilde{\epsilon}_{\mathbf{G}\mathbf{G}'}^{-1}(\mq;0)},
\label{phase}
\een
where $\Omega^2_{\mathbf{G}\mathbf{G}'}(\mathbf{q})$ is given by Eq. \eqref{gpp} and we have changed the $\epsilon$ in the original expression by $\tilde{\epsilon}$ in Eq. \eqref{phase} above, defined at the end of Sec. \ref{sec:embed}. Note that in Eq. \eqref{phase}, although $\omega_p^2$ and $\rho(\mg-\mg')/\rho(\mathbf{0})$ in the $\Omega^2_{\mathbf{G}\mathbf{G}'}(\mathbf{q})$ expression are from the interface as we discussed above, the $\tilde{\epsilon}_{\mathbf{G}\mathbf{G}'}^{-1}(\mq;0)$ from the embedded adsorbate is not necessarily identical to $\epsilon_{\mathbf{G}\mathbf{G}'}^{-1}(\mq;0)$ that is directly calculated for the interface. This introduces discrepancy in the GPP mode frequency $\tilde{\omega}^2_{\mathbf{G}\mathbf{G}'}(\mathbf{q})$ and therefore the poles in the embedding calculation are not exactly at the same positions as those in the interface. However, because $\epsilon_{\mathbf{G}\mathbf{G}'}^{-1}(\mq;0)$ is typically much less than 1 for small $\mg=\mg'$, especially for the systems studied here, we expect the numerical error is not significant. We assess the performance of this approach using numerical results below, and leave further detailed studies on this approximation and its consequences as future work.

\section{Results}
\label{sec:res}

\subsection{The Systems and Computational Details}
\label{sec:comp}

\begin{figure}[htbp]
\begin{center}
\includegraphics[width=3.5in]{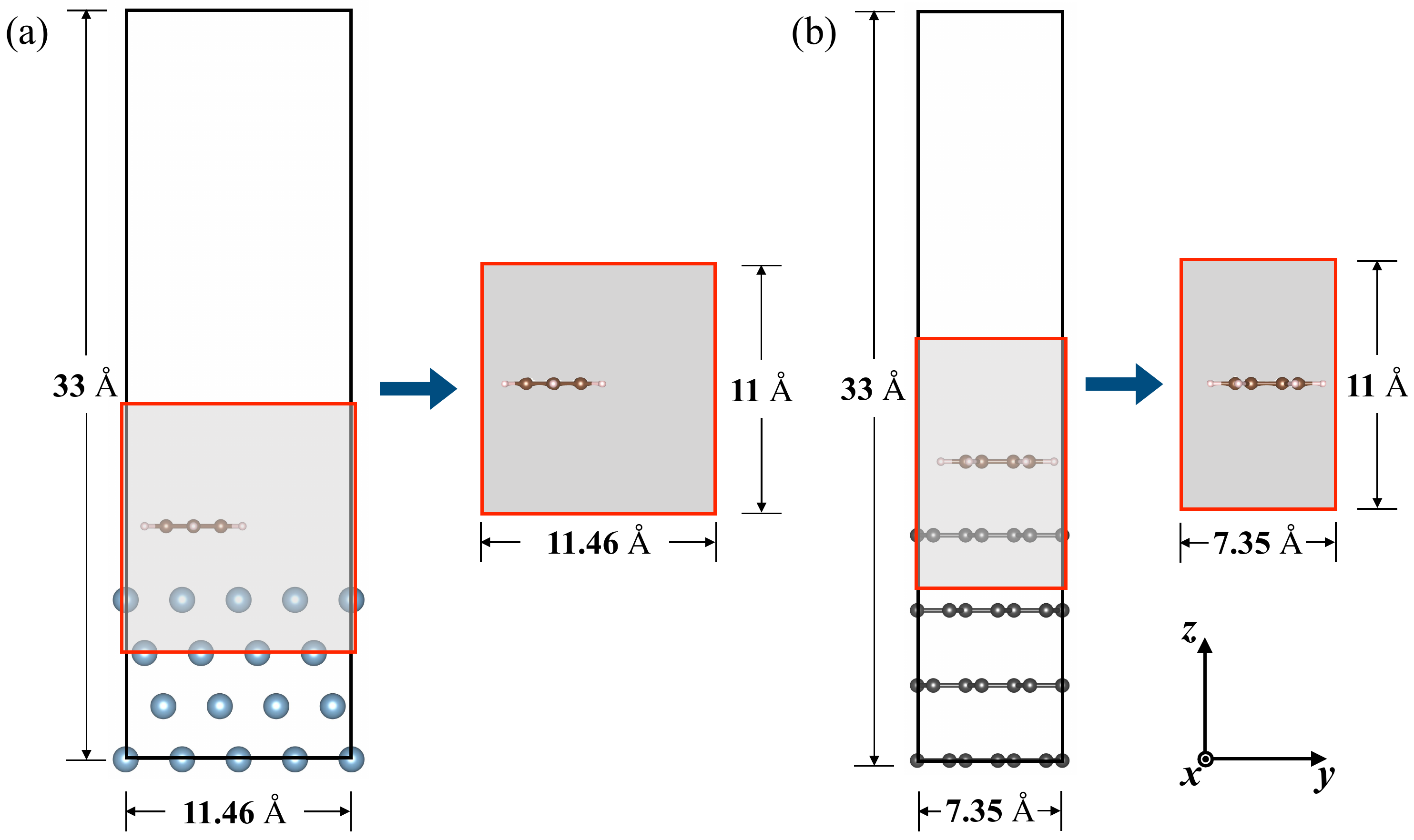}
\caption{(a) Benzene adsorbed on Al (111) surface. (b) Benzene adsorbed on graphite (0001) surface. In both cases, The benzene molecule is sitting flat at 3.24 \AA~ above the surface, and we embed the dielectric environment of the substrate into a periodic simulation cell (the gray area with a red border) containing just the adsorbate atoms, with the $xy$ dimensions being the same as the interface and the $z$ dimension being 1/3 of that of the interface, centered at the molecule. For the purpose of comparison, we choose an embedding region that has the same length along $z$ for the two interfaces.}
\label{system}
\end{center}
\end{figure}

As a proof of principle, we demonstrate our approach using two experimentally and theoretically well-studied weakly coupled molecule-metal interfaces, benzene adsorbed on the Al (111) surface \cite{DMBN00} and benzene adsorbed on the graphite (0001) surface \cite{GT86}. We choose these systems because results from prior $GW$ calculations are available \cite{LJLN19,NHL06}, which are helpful in assessing our approach compared to other $GW$-based approximations. Additionally, the adsorbate and the adsorption height are the same for both interfaces and the difference in results could imply the difference in the effect of the substrate. The systems are shown in Fig. \ref{system}.

For the first system, a benzene molecule is adsorbed at 3.24~\AA~above the Al (111) surface, and the Al substrate is modeled by 4 atomic layers with each layer containing 4$\times$4 Al atoms. A $4\times4\times1$ $\mathbf{k}$-mesh is used. The geometry is taken from Ref. \citenum{LERK17}. For the second system, a benzene molecule is adsorbed at 3.24~\AA~above the graphite (0001) surface, and the graphite substrate is modeled by 4 atomic layers with each layer containing $3\times3$ graphite unit cells (18 carbon atoms). A $6\times6\times1$ $\mathbf{k}$-mesh is used. The geometry is taken from Ref. \citenum{NHL06}. For both systems, periodic boundary conditions are used in all directions, and the size of the interface cell along the $z$ direction is 33~\AA, with about 20~\AA~vacuum. The atomic coordinates of the molecule are fully relaxed, while the substrate atoms are kept fixed in their bulk geometry, so that the reciprocal-space folding for $\chi^0_{\rm sub}$ is exact. A kinetic cutoff of 50 Ry is used in the DFT calculations of both systems, which are performed in \texttt{Quantum ESPRESSO} \cite{QE2017}, using the Perdew-Burke-Ernzerhof (PBE) functional \cite{PBE96}. All $GW$ calculations are performed in \texttt{BerkeleyGW} \cite{BerkeleyGW}, at the $G_0W_0$ level based on the PBE starting point.

The $GW$ calculation starts with the substrate unit cell [Fig. \ref{cartoon}(d)]. The unit cell of the Al (111) surface is modeled by 4 layers with 1 atom on each layer and a $16\times16\times1$ $\mathbf{k}$-mesh. The unit cell of the graphite (0001) surface is modeled by 4 layers with 2 carbon atoms on each layer and a $18\times18\times1$ $\mathbf{k}$-mesh. A dielectric cutoff of 5 Ry (10 Ry) is used in the polarizability calculation of the Al (graphite) surface. This dielectric cutoff we use corresponds to about 5000 bands in the interface systems. We use a Coulomb truncation \cite{I06} along the $z$ direction in the polarizability and self-energy calculations. After that, the reciprocal-space folding to obtain $\bar{\chi}^0_{\rm sub}$ in the substrate supercell [Fig. \ref{cartoon}(e)] follows the procedure described in Ref. \citenum{LJLN19}. Then, $\bar{\chi}^0_{\rm sub}$ is truncated in real space to obtain $\tilde{\chi}^0_{\rm sub}$ in Fig. \ref{cartoon}(b), following what we discussed in Sec. \ref{sec:embed}. For the systems we studied in this work, $\bar{\chi}^0_{\rm sub}$ is truncated into a cell that extends to 5.5~\AA~above/below the adsorbed molecule, with the same $xy$ dimensions as the interface system (see Fig. \ref{system}). For the purpose of comparison, we choose an embedding region that has the same length along $z$ for the two interfaces. After the real-space truncation, we combine the truncated $\tilde{\chi}^0_{\rm sub}$ and the $\chi^0_{\rm mol}$ of the isolated periodic molecular layer, with the latter calculated using the standard RPA approach in \texttt{BerkeleyGW} \cite{BerkeleyGW}. The result is $\tilde{\chi}^0_{\rm tot}$ as we show in Fig. \ref{cartoon}(a'). Finally, the $\tilde{\chi}^0_{\rm tot}$ is inverted to obtain $\tilde{\epsilon}^{-1}$ in this embedded small simulation cell, followed by self-energy calculations. We emphasize that this small adsorbate simulation cell uses the same periodic boundary conditions and the slab Coulomb truncation \cite{I06} as the interface.

\subsection{Comparison of Results from Different Levels of Approximation}

In this section, we compare the $GW$ corrections to PBE level alignments, $\Delta E_{\rm HOMO}$ and $\Delta E_{\rm LUMO}$ (HOMO: highest occupied molecular orbital; LUMO: lowest unoccupied molecular orbital), for the two interfaces we study. The $GW$-predicted HOMO (LUMO) level alignment is simply the HOMO (LUMO) peak position in the PBE PDOS shifted down (up) by $\Delta E_{\rm HOMO}$ ($\Delta E_{\rm LUMO}$). Based on Eq. \eqref{eq1}, we compare these quantities calculated from four levels of $GW$-based approximation: (A) $GW$ self-energy calculation for the interface orbitals that correspond to the molecular resonances, $\braket{\phi_i^{\rm tot}|\Sigma[G^{\rm tot}W^{\rm tot}]|\phi_i^{\rm tot}}$; (B) molecular projection $GW$, $\braket{\phi_i^{\rm mol}|\Sigma[G^{\rm tot}W^{\rm tot}]|\phi_i^{\rm mol}}$, as developed in Refs. \citenum{TDQB11,CTQ17}; (C) dielectric embedding $GW$ developed in this work, $\braket{\phi_i^{\rm mol}|\Sigma[G^{\rm mol}\tilde{W}^{\rm tot}]|\phi_i^{\rm mol}}$, with all quantities calculated in the adsorbate simulation cell [Fig. \ref{cartoon}(a')] that is much smaller in $z$ than the interface cell. As a direct comparison to (C), we perform another dielectric embedding $GW$ calculation (D), but with all quantities calculated in the interface cell, i.e., without the real-space truncation of the supercell substrate polarizability $\bar{\chi}^0_{\rm sub}$. The periodic molecular layer is then placed in the same simulation cell as the interface. In other words, the gray area in Fig. \ref{system} spans the entire interface simulation cell. The difference between (C) and (D) is then solely due to the real-space truncation of $\bar{\chi}^0_{\rm sub}$. The difference between (B) and (D) is due to the screened Coulomb interactions between adsorbate orbitals and substrate orbitals. We show the results in Table \ref{tab:results} and describe details of each method below.

\begin{table}
\caption{Comparison of results from different levels of approximation. Shown are $\Delta E_{\rm HOMO}$ and $\Delta E_{\rm LUMO}$ in eV, the $GW$ corrections to PBE level alignments. $\Delta$gap is the gap change in $GW$ compared to PBE, i.e., $\Delta$gap$=\Delta E_{\rm HOMO}+\Delta E_{\rm LUMO}$. Method (A): $GW$ self-energy calculation for the interface, $\braket{\phi_i^{\rm tot}|\Sigma[G^{\rm tot}W^{\rm tot}]|\phi_i^{\rm tot}}$. Method (B): molecular projection $GW$, $\braket{\phi_i^{\rm mol}|\Sigma[G^{\rm tot}W^{\rm tot}]|\phi_i^{\rm mol}}$, as developed in Refs. \citenum{TDQB11,CTQ17}. Method (C): dielectric embedding $GW$ developed in this work, $\braket{\phi_i^{\rm mol}|\Sigma[G^{\rm mol}\tilde{W}^{\rm tot}]|\phi_i^{\rm mol}}$, with all quantities calculated in the adsorbate simulation cell [Fig. \ref{cartoon}(a')] that is much smaller in $z$ than the interface cell. Method (D): same as (C), but with all quantities calculated in the interface cell without the real-space truncation of substrate polarizability. All $GW$ calculations use the Hybertsen-Louie GPP model \cite{HL86}.}
\begin{center}
\begin{tabular}{c|ccc|ccc}
\hline\hline
\multirow{2}{*}{Method} & \multicolumn{3}{c|}{Benzene on Al (111)} & \multicolumn{3}{c}{Benzene on graphite (0001)} \\
 & $\Delta E_{\rm HOMO}$ & $\Delta E_{\rm LUMO}$ & $\Delta$gap & $\Delta E_{\rm HOMO}$ & $\Delta E_{\rm LUMO}$ & $\Delta$ gap \\
\hline
(A) & 0.72 & 0.36 & 1.08 & 1.34 & 0.72 & 2.06 \\
(B) & 0.74 & 0.99 & 1.73 & 1.40 & 0.39 & 1.79 \\
(C) & 0.99 & 0.90 & 1.89 & 1.61 & 0.04 & 1.65 \\
(D) & 1.01 & 1.05 & 2.06 & 1.51 & 0.15 & 1.66 \\
\hline\hline
\end{tabular}
\end{center}
\label{tab:results}
\end{table}

For method (A), direct $GW$ self-energy calculation of the interface system, we first find the interface orbital that most resembles the gas-phase adsorbate HOMO/LUMO. To do that, we expand the gas-phase adsorbate HOMO/LUMO using the interface orbitals as a basis, at the $\Gamma$ point: $\ket{\phi_\mu^{\rm mol}}=\sum_i \ket{\phi_i^{\rm tot}}\braket{\phi_i^{\rm tot}|\phi_\mu^{\rm mol}}$, where $\mu$=HOMO or LUMO of the isolated periodic molecular layer at the $\Gamma$ point, calculated in the same simulation cell as the interface. Then we identify the $\ket{\phi_i^{\rm tot}}$ with the largest $|\braket{\phi_i^{\rm tot}|\phi_\mu^{\rm mol}}|^2$ as the interface orbital that most resembles the adsorbate HOMO/LUMO, i.e., the molecular resonance. Its energy is very close to the PDOS peak and the difference is mainly due to the $\mathbf{k}$-averaging in PDOS calculations. Then we carry out $GW$ self-energy calculations for these interface orbitals, using $\braket{\phi_i^{\rm tot}|\Sigma[G^{\rm tot}W^{\rm tot}]|\phi_i^{\rm tot}}$. The self-energy corrections to the PBE eigenvalues of $\ket{\phi_i^{\rm tot}}$ are reported in Table \ref{tab:results}. For weakly coupled molecule-metal interfaces, there is often one $\ket{\phi_i^{\rm tot}}$ whose expansion coefficient is significantly larger than others, so the molecular resonance is well-defined.

For method (B), the molecular projection $GW$, we follow the procedure outlined in Refs. \citenum{TDQB11,CTQ17}. For completeness and the purpose of comparison, we briefly describe the approach here. First we compute the expectation value $\braket{\phi_i^{\rm mol}|H^{\rm tot}|\phi_i^{\rm mol}}$, where $i$=HOMO or LUMO of the isolated periodic molecular layer at the $\Gamma$ point. $H^{\rm tot}$ is the KS Hamiltonian of the interface system. We use this value as the mean-field starting point for subsequent $GW$ self-energy calculation, instead of the PBE eigenvalue of $\ket{\phi_i^{\rm tot}}$ as in method (A). Similarly, the exchange-correlation (XC) contribution involved in the $GW$ self-energy calculation is evaluated using $\braket{\phi_i^{\rm mol}|V_{\rm xc}^{\rm tot}|\phi_i^{\rm mol}}$, where $V_{\rm xc}^{\rm tot}$ is the XC potential of the interface system. Although the same $G^{\rm tot}$ and $W^{\rm tot}$ are used here as in method (A), the expectation value of $\Sigma$ on the HOMO/LUMO of the isolated molecular layer is calculated: $\braket{\phi_i^{\rm mol}|\Sigma[G^{\rm tot}W^{\rm tot}]|\phi_i^{\rm mol}}$. This difference in the orbital for which the expectation value of $\Sigma$ is calculated, as well as the difference in the mean-field energy and XC energy, contribute to the difference in self-energy corrections from those calculated using method (A), which we report in Table \ref{tab:results}. Note that when the GPP \cite{HL86} model is used in the self-energy calculation, the plasmon poles are calculated based on the interface system, consistent with $G^{\rm tot}$. 

Before we continue the discussion of embedding $GW$, we would like to point out that there seems to be no definitive conclusion whether method (A) or method (B) should be considered ``the benchmark $G_0W_0$'' for the molecular frontier resonances in a generic molecule-metal interface. This is because both (A) and (B) involve approximations that are of similar quality: method (A) assumes that the mean-field interface orbital is a good approximation to the quasiparticle orbital of the molecular resonance, hence uses $\ket{\phi_i^{\rm tot}}$ as the mean-field starting point for $G_0W_0$; method (B) assumes that the mean-field isolated adsorbate orbital is a good approximation to the quasiparticle orbital of the molecular resonance, hence uses $\braket{\phi_i^{\rm mol}|H^{\rm tot}|\phi_i^{\rm mol}}$ as the mean-field starting point and calculates the expectation of $\Sigma$ on $\ket{\phi_i^{\rm mol}}$. Therefore, whether (A) or (B) leads to better agreement with experiment is in fact system-dependent. From Table \ref{tab:results}, we can see that (A) and (B) generally agree on HOMO of benzene, but not on LUMO, for both interfaces. This is physically because in both interfaces, HOMO is a better defined resonance, characterized by a narrower peak in PDOS than LUMO. We also note that method (B) should be the proper benchmark for comparison of the dielectric embedding $GW$ approach that we develop in this work, because they make the same assumption and calculate the same expectation value of $\Sigma$ on $\ket{\phi_i^{\rm mol}}$ [see Eq. \eqref{eq1}]. 

For the dielectric embedding $GW$, method (C) and method (D), we follow the procedure as described in Sec. \ref{sec:metho} and Sec. \ref{sec:comp}. Since the only difference here is whether or not the real-space truncation of substrate $\bar{\chi}^0_{\rm sub}$ is performed, the difference in results reflects the effectiveness of the truncation. From Table \ref{tab:results} we can see that the truncation only results in minor differences on the order of 0.1 eV based on the GPP model. Additionally, we note that the error is largely from the difference in the plasmon poles: one way to see this is through the results obtained using COHSEX \cite{JDSC14}, where the effect of the plasma frequency is absent since the static dielectric function is used. Such a comparison is shown in Table \ref{tab:cohsex}. The GPP counterpart of first (second) row of Table \ref{tab:cohsex} is simply method (C) [(D)] in Table \ref{tab:results}. We note that for the Al (111) surface, this truncation does not introduce any error, but for the graphite (0001) surface, the error is less than 0.1 eV. This small error is possibly due to the layered nature of graphite, and the truncation is not deep enough in real space (see Fig. \ref{system}).

\begin{table}
\caption{Comparison of dielectric embedding $GW$ results, calculated using COHSEX  with the real-space truncation of $\bar{\chi}^0_{\rm sub}$ (in a small adsorbate cell) and without the real-space truncation of $\bar{\chi}^0_{\rm sub}$ (in the large interface cell). Shown are $\Delta E_{\rm HOMO}$ and $\Delta E_{\rm LUMO}$ in eV. Note that COHSEX shifts LUMO of benzene on graphite downwards (unphysical) instead of upwards.}
\begin{center}
\begin{tabular}{c|cc|cc}
\hline\hline
real-space $\bar{\chi}^0_{\rm sub}$ & \multicolumn{2}{c|}{Benzene on Al} & \multicolumn{2}{c}{Benzene on graphite } \\
truncation & $\Delta E_{\rm HOMO}$ & $\Delta E_{\rm LUMO}$ & $\Delta E_{\rm HOMO}$ & $\Delta E_{\rm LUMO}$ \\
\hline
Yes & 2.55 & 0.18 & 3.44 & -0.64 \\
No & 2.55 & 0.18 & 3.52 & -0.52 \\
\hline\hline
\end{tabular}
\end{center}
\label{tab:cohsex}
\end{table}

For a comparison of all the four methods we used in Table \ref{tab:results}, one can see that in general the dielectric embedding methods [(C) and (D)] are in agreement with the molecular projection approach [method (B)] for all cases considered, with about 0.2-0.3 eV error. Note that experimental techniques to measure level alignments, such as the (inverse) photoemission spectroscopy, have error bars and sometimes large widths in the peak. Therefore, for the purpose of verifying, explaining, and predicting level alignment at molecule-metal interfaces in comparison with experiments, we believe that the dielectric embedding approach presented here is reliable and useful. This is especially the case when considering that the computational cost of such embedding calculations is only a fraction of that of direct $GW$ [method (A)] or molecular projection $GW$ [method (B)] due to the use of a much smaller simulation cell.

We comment that it is critical to use the interface plasma frequency in the embedding $GW$ calculation, as we discussed in Sec. \ref{sec:plasma}. As a comparison, if we perform dielectric embedding $GW$ calculations based on the GPP model using the plasma frequency of the adsorbate, both the gap and the level alignments are not accurate. This is regardless of whether we perform the real-space truncation of the substrate polarizability or not. For instance, for the benzene on the Al (111) system, using the plasma frequency of the adsorbate, the $\Delta E_{\rm HOMO}$ is 0.57 eV (0.42 eV) with (without) the $\bar{\chi}^0_{\rm sub}$ truncation, and the $\Delta E_{\rm LUMO}$ is 1.01 eV (1.08 eV) with (without) the $\bar{\chi}^0_{\rm sub}$ truncation. These results are in sharp contrast to those reported in Table \ref{tab:results}, especially for HOMO. They are also very different from the direct $GW$ and molecular projection $GW$ results.

\section{Discussions}
\label{sec:dis}

We discuss the advancements and current limitations of our approach, and do so in a manner echoing the three issues that we listed at the beginning of Sec. \ref{sec:metho} for a successful dielectric embedding $GW$ approach. First, we showed using both GPP and COHSEX results that it is possible to embed the substrate dielectric effect into a much smaller simulation cell containing just the adsorbate, which greatly reduces the computational cost. The dielectric embedding is achieved via a real-space truncation of the substrate $\chi^0$, followed by a summation of the truncated substrate $\chi^0$ and the periodic molecular $\chi^0$. For simple metals whose Thomas-Fermi screening length is short, we expect that this truncation works very well, such as the Al (111) case shown in this work. However, in the other limit, for semiconductors in which the dielectric screening decays very slowly as distance, the real-space truncation proposed here might become problematic, and is worth of scrutinization on a case-by-case basis. An effective real-space truncation is needed to successfully extend this approach to semiconductor substrates.

Second, we recognize that full-frequency $GW$ calculations are still computationally challenging for the dielectric embedding. We have verified that a full-frequency treatment in the reciprocal-space folding of substrate $\chi^0$ is indeed practically feasible and straightforward, since it only involves $\mg$-vector matching and scales linearly with respect to the number of frequency grids. It is the real-space truncation of substrate $\chi^0$ that is currently prohibitive for a full-frequency treatment due to its high computational cost. Future developments on algorithms for an efficient real-space truncation are needed to enable full-frequency dielectric embedding. At present, we have to focus on and employ the GPP model for the dielectric function. It is then critical to use the interface plasma frequency in the embedding calculation of the molecular adsorbate, so that the $GW$ self-energy could reflect the physical level alignment at the interface. There is, however, a small caveat here. As we pointed out in the discussion of Eq. \eqref{phase}, we use the interface plasma frequency, but the GPP mode frequency is still not exactly the same as that used in the interface $GW$ calculation, due to the difference in $\epsilon_{\mg\mg'}^{-1}$ in the denominator of Eq. \eqref{phase}. Future work is therefore desired to study the quality of this approximation in more detail. One possible way is to check the extent to which the generalized $f$-sum rule is violated, which is the underlying principle that the GPP model is based on \cite{HL86}. 

Third, because we focus on weakly coupled molecule-metal interfaces in this work, in the self-energy calculations, we have neglected the terms arising from the screened Coulomb interactions between the isolated adsorbate orbitals and interface orbitals that are localized on the substrate. In other words, if we compare the second and the third expressions in Eq. \eqref{eq1}, we have neglected terms like $\int \sum_j \phi_i^{{\rm mol}*}(\mr')\phi_j^{{\rm sub}*}(\mr')\phi_j^{\rm sub}(\mr)\phi_i^{\rm mol}(\mr)W(\mr,\mr';\omega)\,d\mr d\mr'$ in the self-energy calculation in the dielectric embedding $GW$ approach. The numerical difference between results obtained from method (B) and method (D) in Table \ref{tab:results} is a reflection of such contributions. In general, the more a molecular orbital is hybridized with substrate orbitals, the more important these contributions will become. Since $W(\mr,\mr';\omega)$ decays as $|\mr-\mr'|$ increases, we expect that such contributions are only important for the screened Coulomb interaction between adsorbate molecular orbitals and substrate orbitals that are localized near the top surface. In the limit of negligible orbital hybridization upon adsorption (physically, this is the case for interfaces with a large adsorption height), we expect that such contributions vanish and the dielectric embedding becomes exact.

\section{Conclusion}
\label{sec:conc}

Accurate and efficient calculations of quasiparticle properties at heterogeneous interfaces are in general computationally challenging. In this work, we develop a dielectric embedding formalism, to confine the expensive $GW$ calculations within the adsorbate simulation cell that is much smaller in size than the interface simulation cell. This is achieved through a real-space truncation of the substrate polarizability, followed by a summation of the truncated substrate polarizability with the adsorbate polarizability. In order to incorporate the difference in the plasma frequency between the adsorbate and the interface, it is key to use the plasma frequency of the interface in embedding $GW$ calculations of the adsorbate. As a proof of principle, we showed the feasibility and accuracy of our approach using two well-studied weakly coupled molecule-metal interfaces, and compared the results against direct $GW$ calculations of the interface and the molecular projection $GW$ approach. We showed that we can achieve similar accuracy at a much smaller computational cost. We believe that our approach paves the way for future development of embedding formalisms in the framework of many-body perturbation theory, for a broad range of heterogeneous interfaces.

\section{Acknowledgements}

Z.-F. L. thanks Wayne State University for generous start-up funds, the Ebbing faculty development award, and the Grid Computing for computational support. Additional computational support for method development is made possible through the use of the Center for Nanoscale Materials at the Argonne National Laboratory, an Office of Science user facility, which was supported by the U.S. Department of Energy, Office of Science, Office of Basic Energy Sciences, under Contract No. DE-AC02-06CH11357. Large-scale benchmark calculations used computational resources of the National Energy Research Scientific Computing Center (NERSC), a U.S. Department of Energy Office of Science User Facility operated under Contract No. DE-AC02-05CH11231.

\bibliography{lit_GWembed}

\begin{thebibliography}{43}
\bibitem [{not()}]{note1}%
  \BibitemOpen
  \href@noop {} {}\bibinfo {howpublished} {Technically, we create a charge
  density file \texttt{RHO} with the header (containing the
  $\mathbf{G}$-vectors) taken from the adsorbate in the small simulation cell,
  and $\rho(\mathbf{G})$ values taken from the interface in the large
  simulation cell.}\BibitemShut {Stop}%
\end{thebibliography}

\end{document}